\documentclass[submission,copyright,creativecommons,noncommercial,sharealike]{eptcs}
\providecommand{\event}{PERR 2019} 
\usepackage{breakurl}             
\usepackage{underscore}       
\usepackage{graphicx}
\usepackage{amssymb}
\newtheorem{theorem}{Theorem}

\newtheorem{proposition}[theorem]{Proposition}
\newtheorem{example}[theorem]{Example}

\title{On Quantitative Comparison of\\ Chemical Reaction Network Models} 

\author{Ozan Kahramano\u gullar\i
\institute{University of Trento, Department of Mathematics}
\email{ozan-k.com}
}
\def\titlerunning{On Quantitative Comparison of Chemical Reaction Network Models}
\def\authorrunning{O. Kahramano\u gullar\i}
\begin{document}
\maketitle

\begin{abstract}
Chemical reaction networks (CRNs) provide a convenient language 
for modelling a broad variety of biological systems. 
These models are commonly studied with respect to the 
time series they generate in deterministic or stochastic 
simulations. Their dynamic behaviours are 
then analysed, often by using deterministic methods based on  
differential equations with a focus on the steady states.
Here, we propose a method for comparing CRNs 
with respect to their behaviour in stochastic simulations. 
Our method is based on using 
the flux graphs that are delivered by 
stochastic simulations as abstract 
representations of their dynamic behaviour.
This allows us to compare the behaviour of 
any two CRNs for any time interval, and   
define a notion of equivalence on them that overlaps 
with graph isomorphism at the lowest level of representation. 
The similarity 
between the compared CRNs can be quantified in terms of their distance. 
The results can then be used to refine the models 
or to replace a larger model with a smaller one 
that produces the same behaviour or vice versa.
\end{abstract}

\section{Introduction}

Chemical reaction networks (CRNs) constitute an expressive language 
for building models that are dedicated to studying dynamical 
behaviour of biological systems. Many implementations of
simulation languages for systems biology use CRN 
representations for their back-ends that perform deterministic 
or stochastic simulations, e.g., \cite{BFKF18,SF12,LYPEP11}.  
From a modelling, or programming, perspective, 
 the main challenge in the context of 
 the complex make-up of biological systems is
 to come up with simpler models that are effective 
in addressing biological queries. 
In such a setting, it is a common 
practise to compare different models that are built to produce a certain 
behaviour.  The methods for comparison are often devised on a case-by-case basis, and
they typically rely on the deterministic representations of the CRNs. 
The comparisons are commonly performed on the steady states. The notion and the extent 
of acceptable variation in behaviour between compared models depend on
the system and the query in question.

Despite the limited number of formal means, drawing parallels between  
various models of biological systems is central to many investigations in this field.  
Given that existing efforts are often limited by the measurement of ad hoc 
model signals, it is inherently challenging to provide a general method for 
the task. To this end, we propose a methodology for comparing models in 
relation to the dynamic behaviour given by their flux graphs resulting 
from stochastic simulations \cite{flux,Kah17}.  Here, flux is defined in terms of the intensity 
of resource flow during stochastic simulation from one reaction to another. 
The flux graphs are directed graphs  that display how many of  which network
species instances flow between which model components during any chosen time 
interval  of the simulation. 
As this provides a mathematical structure that quantifies the model dynamics, we use this 
information as a discrete abstraction of the model behaviour that can be compared with other 
structures obtained in the same way. The similarity between simulations with CRNs can 
then be quantified by resorting to a distance function, which we show to be a metric.  
In the following, we illustrate our method and discuss different models,
whereby we argue that models separated with a smaller distance 
produce more similar behaviours.

\section{Chemical reaction networks and mass action semantics}

Let $\mathcal{N}$ denote a chemical reaction network (CRN)
consisting of a set of  $N$ reactions defined on $M$ number of species,
and an initial state, which is a vector of the quantities of the M species.  In  $\mathcal{N}$, each reaction  
$$
m_1 R_1 + \dots +m_l  R_l  \stackrel{\rho}{\rightarrow}  n_1 P_1 + \dots + n_r P_r
$$
describes the reactants on the left-hand side that are replaced by the products on the right-hand side with rate $\rho$. 
 The constants $m_1, \ldots, m_l$
and $n_1, \ldots, n_r$ are positive integers that denote the multiplicity of the reactants  
that are consumed and the  products that are produced.
When such a reaction occurs, it modifies the state vector.
That is, if species $R_1,\ldots, R_l$ are present in their multiplicities  $m_1, \ldots, m_l$, they are consumed,
and the species $P_1, \dots, P_r$ are produced, again in their respective multiplicities $n_1, \ldots, n_r$.  
The reaction rate constant $\rho$ is a positive real number, which determines 
how often a reaction occurs in a system. 
According  to the mass action kinetics, the probability of a reaction's firing at a 
particular state is proportional with $\rho$ 
and the number of possible combinations of its reactants at that state. 

\begin{example}
\label{example:lotka}
The Lotka-Volterra  network  $\mathcal{N}_1$  models the interactions of a predator X with a prey Y. 
$$
\{ \,
r_1 : X \stackrel{100.0}{\longrightarrow} \cdot \,, \quad 
r_2 : Y  \stackrel{300.0}{\longrightarrow} 2Y \,,  \quad 
r_3 : X + Y  \stackrel{1.0}{\longrightarrow} 2X \, \}
$$
The network is initiated with 100 X and 100 Y individuals; the initial state is the vector 
$S_{0} =\langle 100, 100\rangle$.  
\end{example}

\begin{example}
\label{example:sir}
The SIR network $\mathcal{N}_2$ models the spread of an infection in a population, whereby 
the individuals infect each other by direct interactions.
An individual who recovers from the illness has immunity.
$$
\{ \, r_1 : S + I  \stackrel{\beta}{\longrightarrow}  I + I,  \quad   \;  r_2 : I  \stackrel{\gamma}{\longrightarrow}  R \;  \}
$$
Each individual can be susceptible (S), infected (I), or resistant (R). 
Thus an initial state 
with 40 susceptible, 1 infected and no resistant is given by the vector $S_{0} =\langle 40, 1, 0\rangle$.  
\end{example}

\begin{figure}[!b]
$$
\begin{array}{c}
\includegraphics[width=56mm]{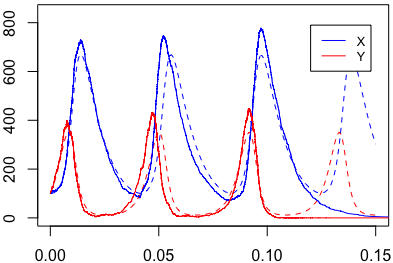}
\qquad \qquad
\includegraphics[width=56mm]{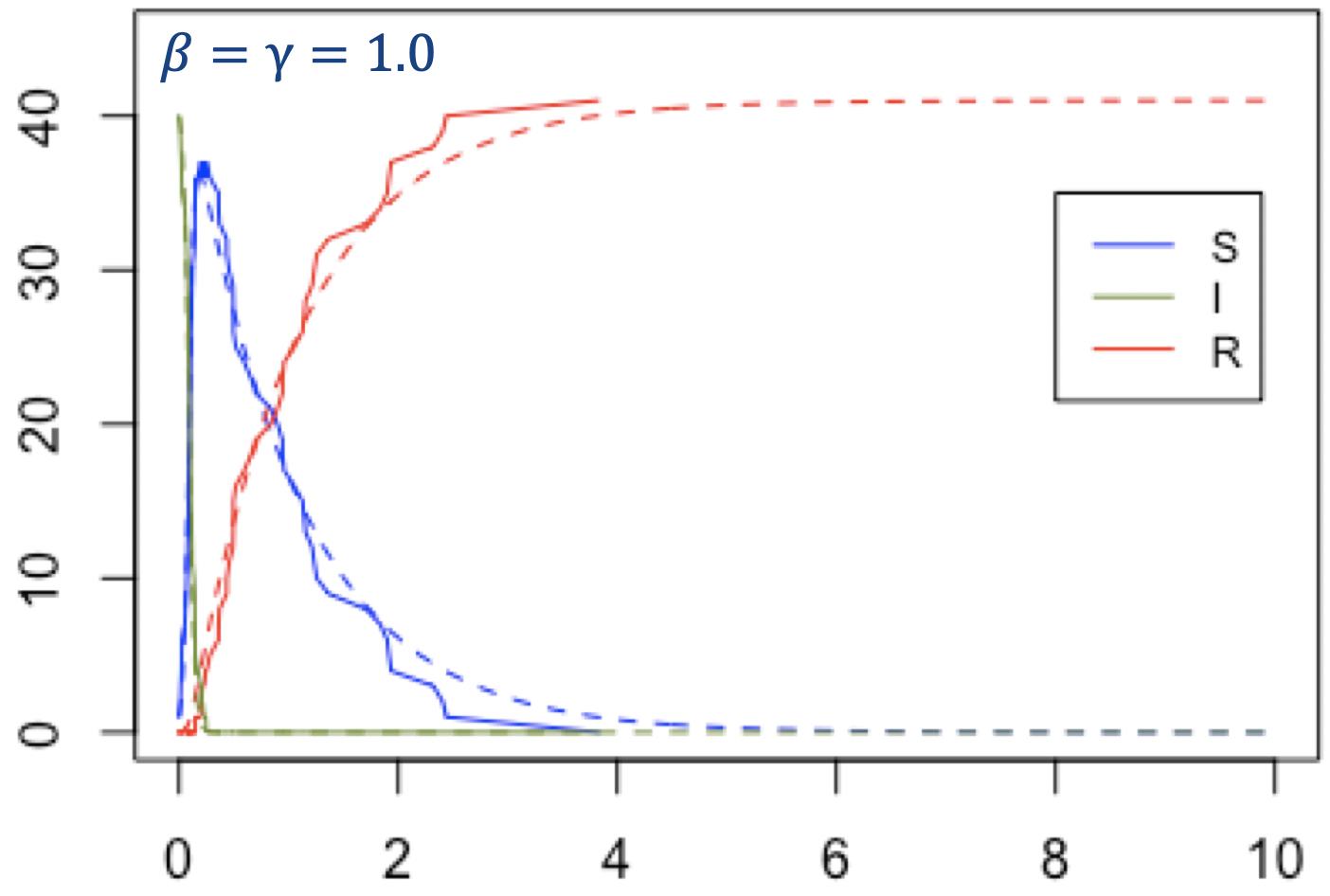}
\end{array}
$$
\caption{Time series of the CRNs in Examples 
\ref{example:lotka} and \ref{example:sir}. The dashed lines are deterministic simulations.}
\label{figure:CRN:examples}
\end{figure}

Stochastic trajectories of CRNs are commonly generated 
by using Gillespie's stochastic simulation algorithm (SSA) \cite{Gil77}. 
These simulations are then commonly depicted as time series. SSA is a 
Monte Carlo algorithm based on mass action with a continuous time 
Markov chain semantics. Thus, the simulations with SSA, at their limit, 
overlap with the deterministic ordinary differential equation simulations. 
However, stochastic simulations reflect the fluctuations in 
species numbers and extinctions that may arise due to low species numbers, 
which are not perceivable in the deterministic setting, e.g., as  in Figure \ref{figure:CRN:examples}.   

\begin{figure}[!t]
$$
\begin{array}{l}
\footnotesize{\verb|reactions|}\\
\footnotesize{\verb|r01: A     ->  B , 1.0;|}\\
\footnotesize{\verb|r02: A     ->  C , 1.0;|}\\
\footnotesize{\verb|r03: B + C ->  D , 1.0;|}\\
\\
\footnotesize{\verb|initial state|}\\
\footnotesize{\verb|100 A|}
\end{array}
\begin{array}{c}
\includegraphics[width=120mm]{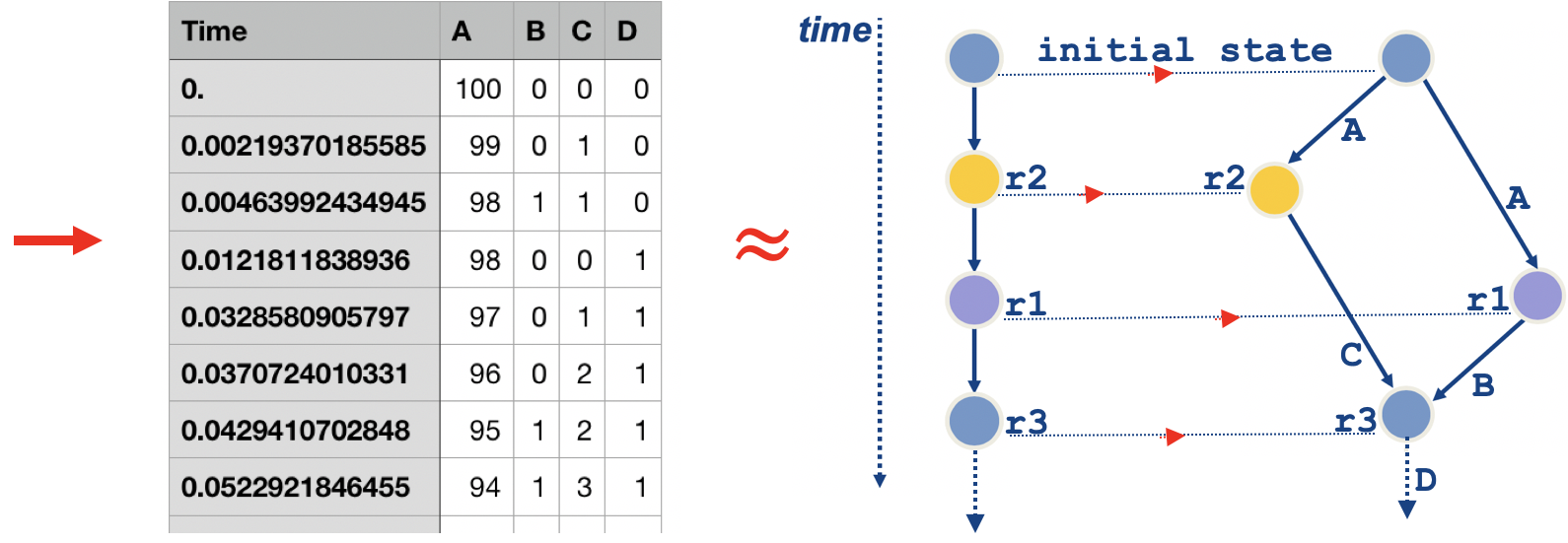}
\end{array}
$$
\caption{The fSSA algorithm \cite{flux,Kah17} extends the stochastic simulation algorithm \cite{Gil77} 
by logging the dependencies of each reaction instance by keeping track of the origin 
of each species that reaction instance consumes. 
This results in a directed acyclic graph that complements the time series data.}
\label{figure:trajectory:to:flux}
\end{figure}

\section{Flux graphs of chemical reaction networks}

A stochastic simulation with a CRN can be considered as a mathematical 
reduction from a complex structure, i.e., the CRN, to a 
simpler structure, i.e., the simulation trajectory. 
As a result of this 
reduction, the algorithm delivers a sequential structure, whereby the 
reaction instances occur one after another starting from the initial state.   
Because of the Markov chain semantics, 
there cannot be any simultaneous reaction instances.
Each reaction instance modifies the state at which it occurs and results 
in a new state. The time stamps of the reaction instances can be seen 
as a manifestation of the sequential total order structure.  
However, a different point of view can be obtained  
if we observe the reaction instances from the point of view of their 
interdependencies due to the resources that they consume and produce: each reaction 
instance consumes resources that were produced by another reaction, 
and produces others that become available for consumption.  
These dependencies result in 
a partial order structure or a directed acyclic graph (dag). 
When we superimpose this dag on to the time series, it becomes possible 
to treat the simulation trajectory as an interleaving of a concurrent computation, 
whereby the dag edges display the resource dependencies. A graphical visualisation
of this idea is depicted in Figure \ref{figure:trajectory:to:flux} on a simple CRN.

\begin{figure}[!b]
$$
\begin{array}{c}
\includegraphics[width=50mm]{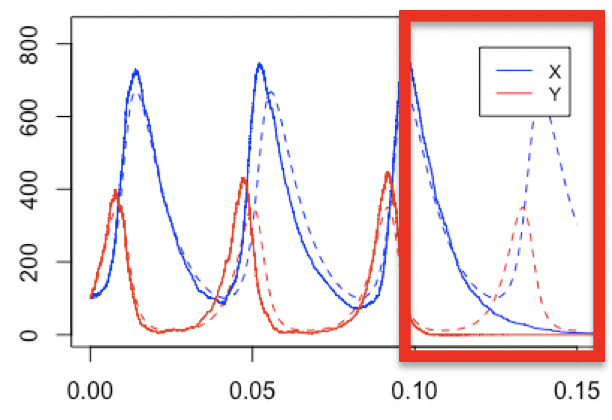}
\end{array}
\begin{array}{c}
\includegraphics[width=8mm]{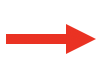}
\end{array}
\begin{array}{c}
\includegraphics[width=34mm]{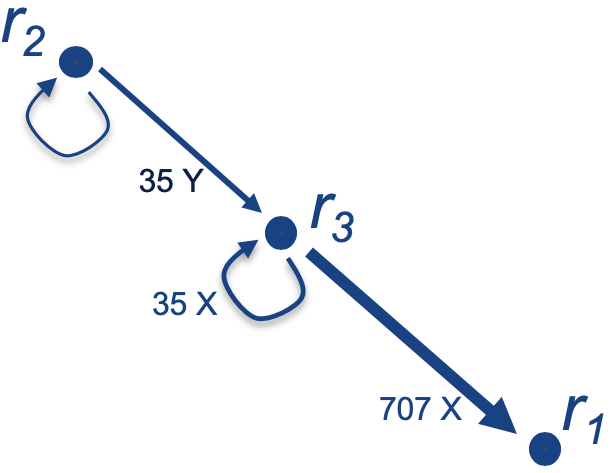}
\end{array}
\begin{array}{c}
\includegraphics[width=1.2mm]{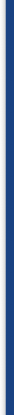}
\end{array}
\begin{array}{c}
\includegraphics[width=48mm]{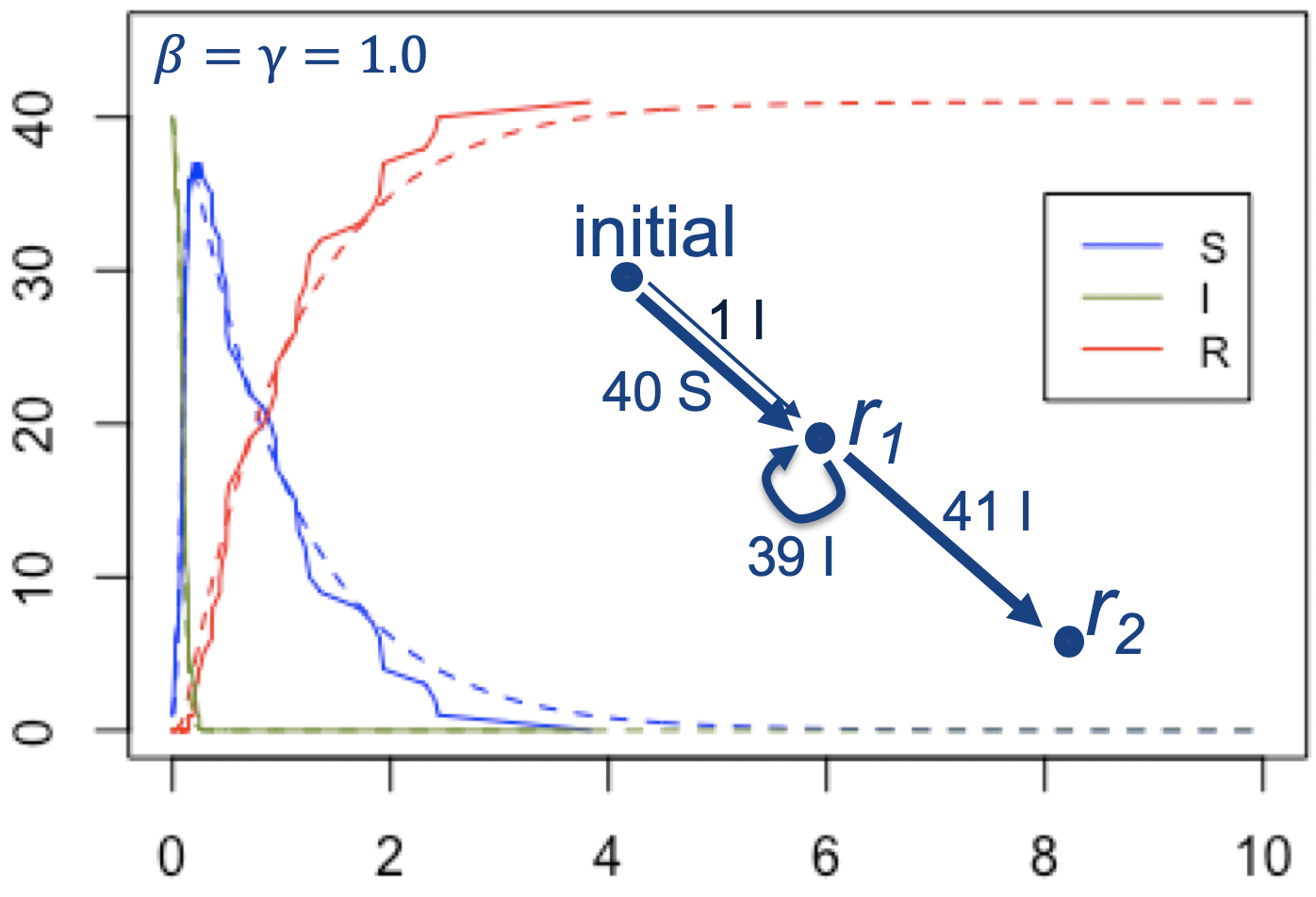}
\end{array}
$$
\caption{Flux graphs of the simulations in Figure \ref{figure:CRN:examples}. 
The graph 
on the left displays the fluxes between the time points 
0.1 and 0.15, where the species go extinct. The graph on the right displays the fluxes 
of the  network in Example \ref{example:sir}  for the complete simulation.}
\label{figure:flux:example} 
\end{figure}

In previous work, we have introduced a conservative extension of 
Gillespie's stochastic simulation algorithm \cite{Gil77}, called fSSA, that generates these graphs. 
In addition to the time series, fSSA logs the dependency 
data by introducing a constant cost to each 
simulation step \cite{flux,Kah17}.  
The algorithm then folds these dags into \emph{flux graphs} that provide a quantification
of the flow of resources between reaction instances for any user-specified time interval.
For the details of the algorithm we refer to \cite{flux,Kah17}. In the following, we describe 
the  structure of the flux graphs, and how they are used for comparing CRNs.   

The flux graphs are edge-coloured 
directed graphs, denoted by $\mathcal{F}[t,t']$, where $t,t'$ denote the time points for 
the beginning and the end of the flux interval. Each edge of the graph is 
of the form $p \stackrel{x,n}{\longrightarrow} q$, where $p$ and $q$ are nodes 
representing the reactions of the CRN, $x$ is a network species, and $n$ is the weight. 
The edge colour $x,n$ on the arrow denotes that between time points 
$t$ and $t'$, species $x$ flowed from $p$ to $q$ with a weight of $n$. 
The weight $n$ denotes the multiplicity of the species $x$ that are logged 
to have flowed from $p$ to $q$.
Figure \ref{figure:flux:example} displays example flux graphs of the 
 simulations in Figure \ref{figure:CRN:examples}.

\section{Comparing chemical reaction networks}
 
The method for comparing CRNs uses the flux graphs as discrete representations 
of the dynamic behaviours of the CRNs. Given two networks to be compared, 
the method takes two parameters.  

The first parameter is the number of simulations, $k$,  to be considered 
for the comparison. 
This parameter 
is used to run $k$ number of simulations, and the mean fluxes of the $k$ 
simulations is  considered for the comparison. A smaller $k$ exposes the 
stochastic  effects and the noise in the system. In contrast, 
a larger $k$ can be used to factor for the stochastic effects.
This way,  it becomes possible to expose the mean behaviour, similar 
to the use of deterministic simulations with ordinary differential equations.
It is also possible to use different $k$ parameters for the two compared networks. 

The  second parameter of the method is the time interval to be considered for the 
comparison, which can be the complete simulation duration or any other 
interval within it. Again, it is possible to use two different intervals  
for the two compared networks. 

\begin{figure}[!b]
$$
\begin{array}{ccc}
\includegraphics[width=45mm]{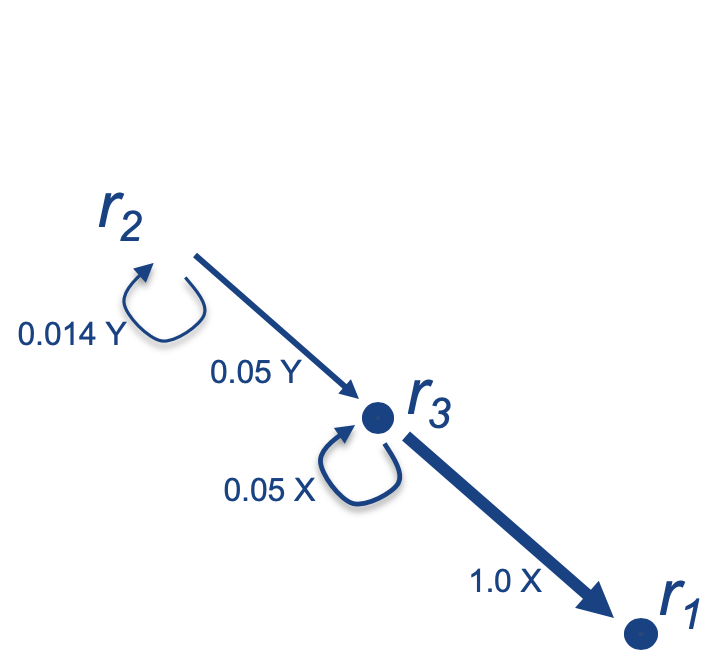}
&
\qquad
\includegraphics[width=45mm]{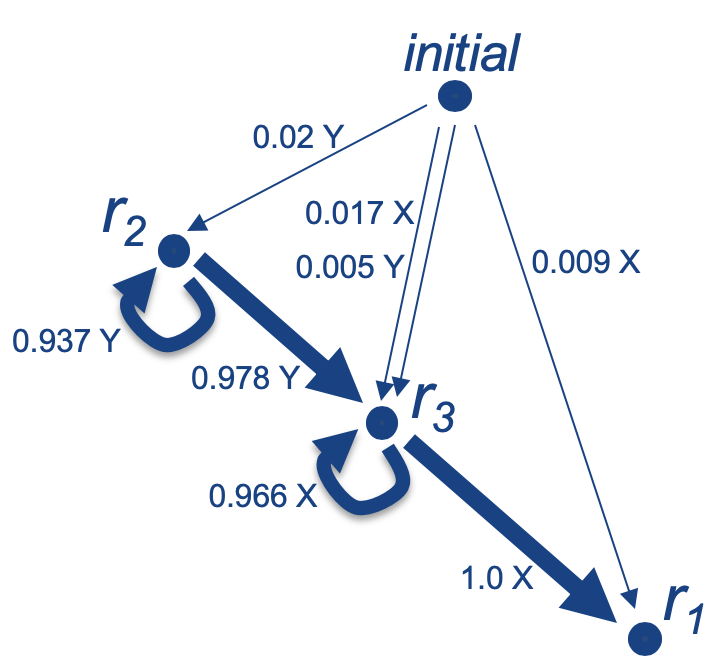}
&
\quad \;\;
\includegraphics[width=45mm]{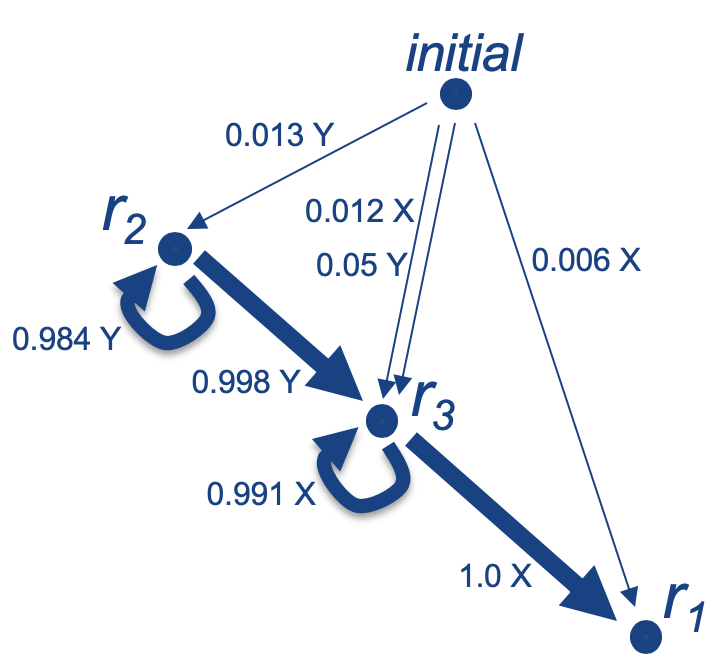}\\[2pt]
k =1, \mathcal{F}[0.1,0.15] & \qquad k =1, \mathcal{F}[0,0.25] &  \qquad k = 10, \mathcal{F}'[0,0.25] 
\end{array}
$$
\caption{A comparison of the CRN in Example \ref{example:lotka} with different time intervals and $k$ parameters.}
\label{example:lotka:window:vs:mean} 
\end{figure}

The method is performed by first running $k$ simulations on the two CRNs, 
and computing the flux graphs of these simulations for the given intervals. 
The comparison is performed on the mean flux graphs of each set of simulations. 
This results in the two flux graphs, one for each network. Each flux graph is then 
normalised by the maximum flux in that network. This way, each weight $n$ of each flux edge 
is replaced with a weight $n/m$, where $m$ denotes the maximum flux value in that network.
Thus, each weight in the flux graph takes a value between $0$ and $1.0$.

\begin{example} 
Let us consider the CRN in Example \ref{example:lotka}. The single run 
of a simulation depicted in Figure \ref{figure:CRN:examples} results in an extinction, 
which is different from the deterministic behaviour of the system. 
A comparison of the normalised flux behaviour of the complete simulation 
with the extinction window in the time interval between $0.1$ and $0.15$ 
in Figure \ref{figure:flux:example} 
reflects this contrast. 
As depicted in Figure \ref{example:lotka:window:vs:mean}, 
this contrast becomes even more amplified 
when  we observe the mean flux behaviour of 10 simulations, 
which is closer to the deterministic setting. In particular,
the normalised mean fluxes display a more balanced distribution between reactions 
in comparison to the fluxes in a single simulation that results in an extinction.
\end{example}

\begin{figure}[!b]
$$
\begin{array}{l}
\begin{array}{c}
\rotatebox{90}{$r_1 \rightarrow r_1$}
\end{array}
\begin{array}{ccc}
\includegraphics[width=50mm]{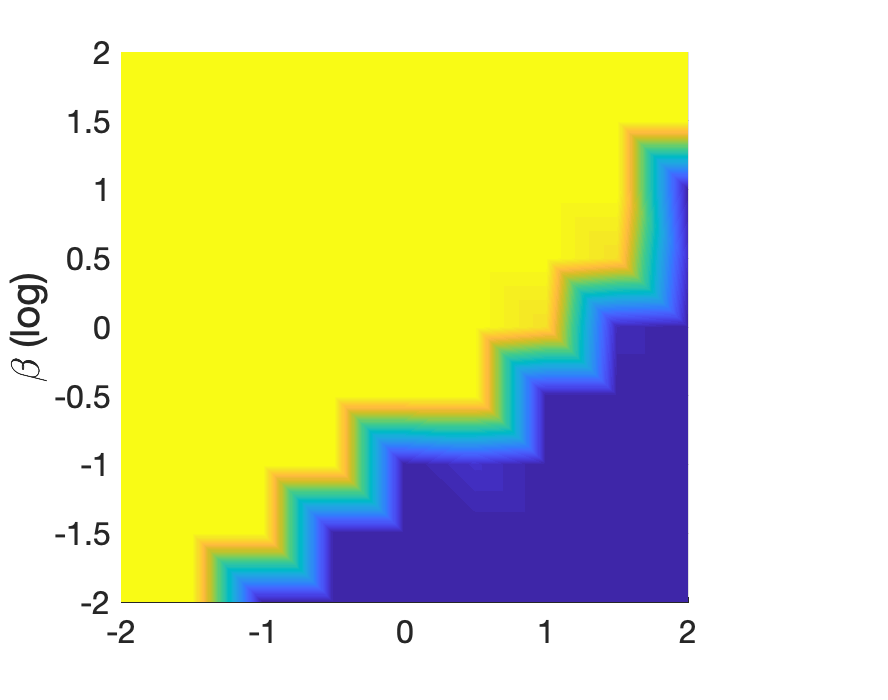}
&
\includegraphics[width=50mm]{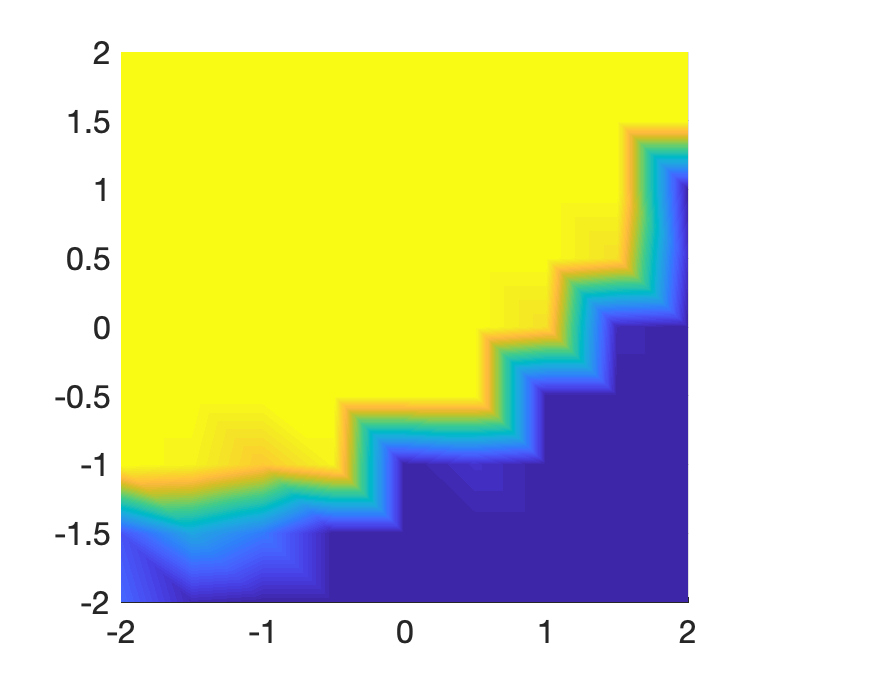}
&
\includegraphics[width=50mm]{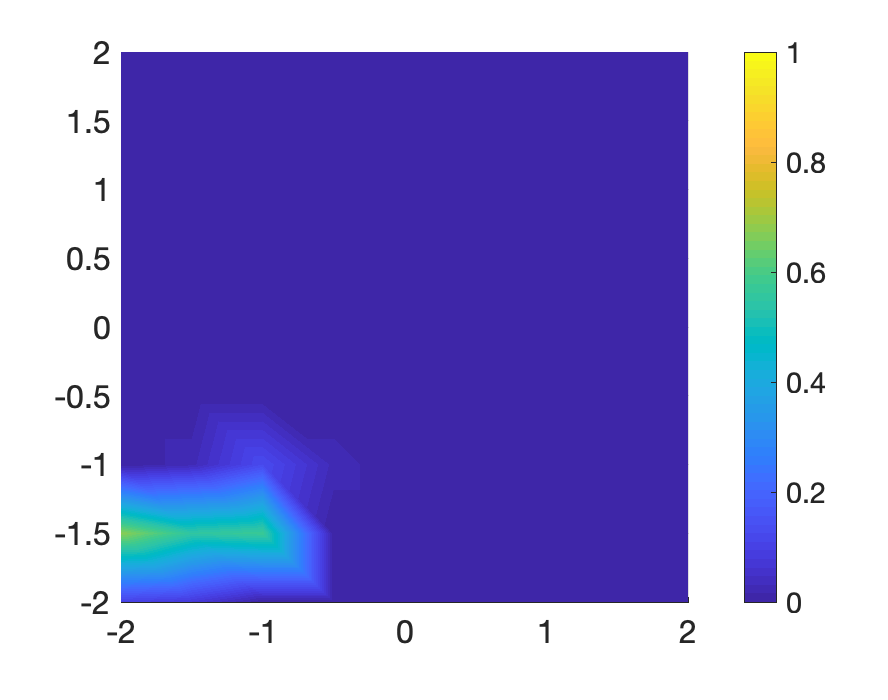}
\end{array}
\\
\begin{array}{c}
\rotatebox{90}{$\quad \; r_1 \rightarrow r_2$}
\end{array}
\begin{array}{ccc}
\includegraphics[width=50mm]{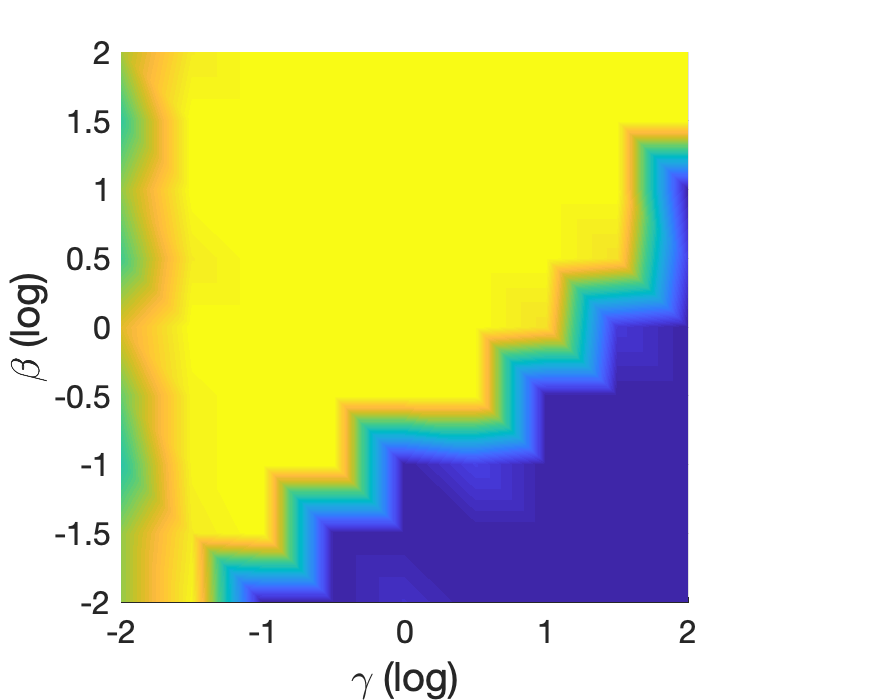}
&
\includegraphics[width=50mm]{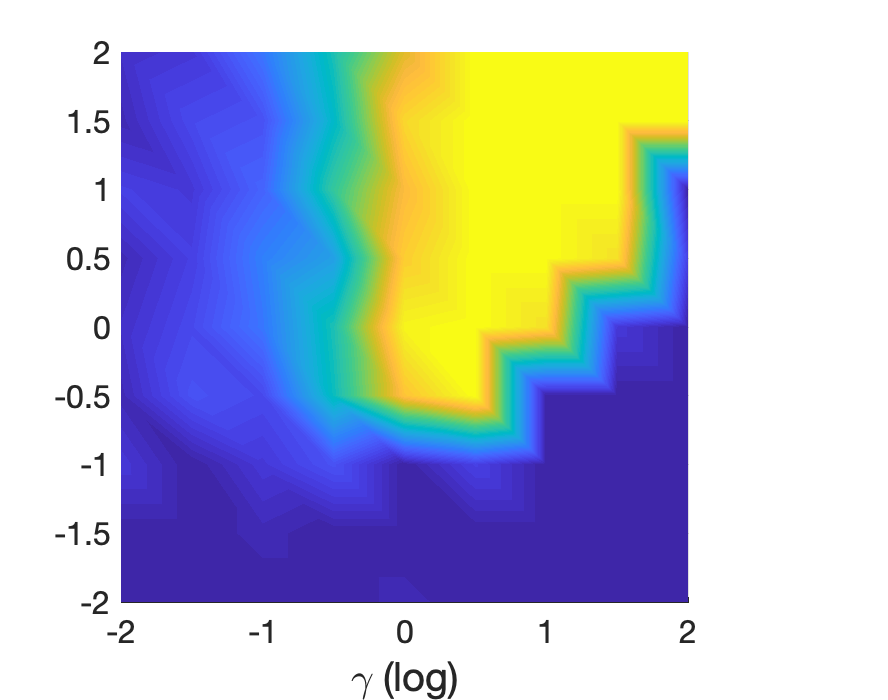}
&
\includegraphics[width=50mm]{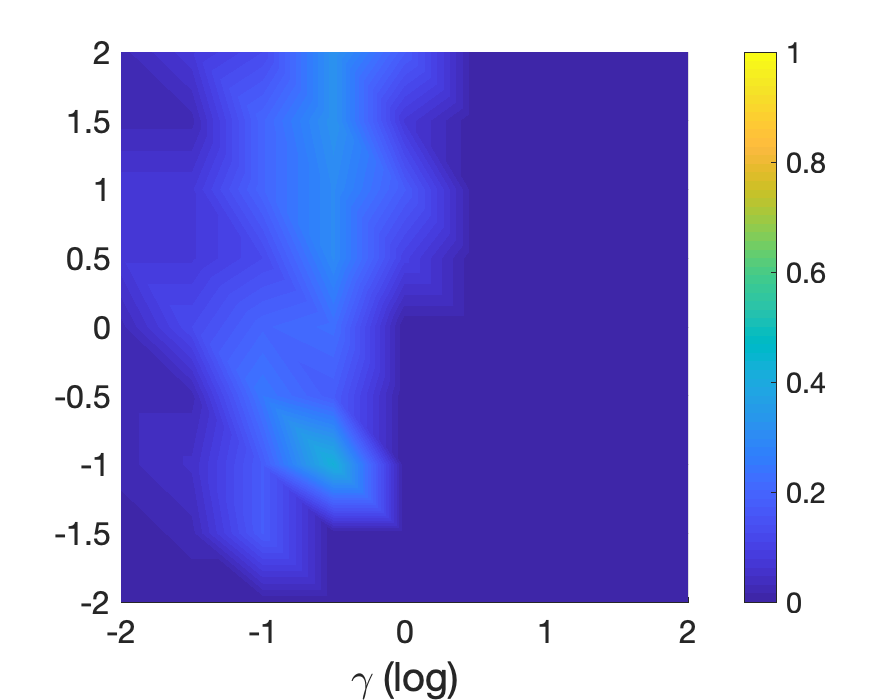}\\[2pt]
\;\; k = 1, \mathcal{F}[0,100] \quad
&
\;\; k = 1,\mathcal{F}[0,2] \quad
&
\;\; k =1, \mathcal{F}[2,4] \quad 
\end{array}
\end{array}
$$
\caption{The mesh of the normalised fluxes resulting from 81 simulations in different time intervals, 
where 
the rates $\beta$ and $\gamma$ of $\mathcal{N}_1$ in Example \ref{example:sir} 
are scanned 4 orders of magnitude on the axes in logarithmic scale. The top row displays 
the  $r_1 \stackrel{I}{\rightarrow} r_1$ fluxes and the bottom row displays the 
$r_1 \stackrel{I}{\rightarrow} r_2$ fluxes.}
\label{figure:sir:scatter:plot}
\label{figure} 
\end{figure}

\begin{example}[sensitivity analysis]
\label{example:sensitivity:anlaysis}
The SIR network $\mathcal{N}_2$ has the rate parameters $\gamma$ and $\beta$ 
that are instantiated as $1.0$ in the stochastic simulation depicted in 
Figure \ref{figure:CRN:examples}. 
The flux graph in Figure \ref{figure:flux:example} together 
with the time-series plot shows that all the susceptible individuals become 
infected and they recover before the time point $4.0$ is reached.
To observe how the parameters $\gamma$ and $\beta$  
influence the dynamic behaviour, we have
performed a set of simulations by 
varying these two parameters within a spectrum of 4 orders of magnitude 
from $10^{-2}$ to $10^{2}$ in the logarithmic scale. 
This has resulted in $9 \times 9 = 81$ simulations.
We have then visualised the effect of these
variations on the two flux edges of this network for the interval from time $0$ to $2$, 
and compared these with the fluxes of the complete simulation. 
The process, whereby a previously infected individual infects another individual, is 
given by the flux edge  $r_1 \stackrel{I}{\rightarrow} r_1$. The recovery of 
infected individuals is given by the fluxes to the reaction $r_2$, that is, the 
flux edge  $r_1 \stackrel{I}{\rightarrow} r_2$. Figure \ref{figure:sir:scatter:plot} 
displays the normalised flux values in these simulations for the complete 
simulation as well as for the time intervals 0 to 2 and 2 to 4.   As it can be 
seen in these plots, the fluxes in the shorter intervals 
($\mathcal{F}[0,2]$ and $\mathcal{F}[2,4]$) 
add up to contribute 
to the fluxes of the complete simulation ($\mathcal{F}[0,100]$).  
While a speed up in $\alpha$ results in the infection of the whole population, 
a smaller $\gamma$ delays the recovery.
\end{example}

The similarity between the two normalised flux graphs $\mathcal{F}$ and $\mathcal{F}'$ 
is determined by a distance function function.  
The distance between the two flux graphs $\mathcal{F}$ and $\mathcal{F}'$ is given by the sum 
of  squared differences of the edge weights with each colour (species). If an edge 
with a  colour does not exist, its weight is assigned a value of $0$. 
The distance  function $\delta$ is formally defined as follows:
$$
\delta(\mathcal{F}_1,\mathcal{F}_2)  = 
\mathsf{sqrt} \big( \;\;\;
\sum_{p \stackrel{x}{\longrightarrow} q \; \in \;\mathcal{F}_1 \cup \mathcal{F}_2}
\!\!\!\!\!\!
   (w_{1} - w_{2})^2  \;\;\textrm{ such that for } i = 1,2 ,  \;\;
  w_i = \left\{
 \begin{array}{ll}
 n_i , &  p \stackrel{x,n_i}{\longrightarrow} q \in \mathcal{F}_i\\[2pt]
0  , & \textit{otherwise}
\end{array}
\right.
\;\;\big)
$$

\begin{proposition}
The distance function $\delta$ is a metric. That is, for any flux graph
$\mathcal{F}_1$, $\mathcal{F}_2$ and $\mathcal{F}_3$:
\begin{enumerate}
\item
$\delta(\mathcal{F}_1,\mathcal{F}_2) \geq 0$;
\item
 $\delta(\mathcal{F}_1,\mathcal{F}_2) = 0\;$ if and only if $\;\mathcal{F}_1 = \mathcal{F}_2$;
\item
$\delta(\mathcal{F}_1,\mathcal{F}_2) =\delta(\mathcal{F}_2,\mathcal{F}_1)$; 
\item
$\delta(\mathcal{F}_1,\mathcal{F}_3) \leq  
\delta(\mathcal{F}_1,\mathcal{F}_2) + 
\delta(\mathcal{F}_2,\mathcal{F}_3)$.  
\end{enumerate} 
\end{proposition}
\noindent \textit{Proof:}
The function $\delta$ can be mapped to Euclidean distance by mapping 
flux graphs to Euclidean vectors as follows: 
for the flux graphs $\mathcal{F}_1$, $\mathcal{F}_2$ and $\mathcal{F}_3$, let 
$E = \{  (p,x,q) \; | \; (p, (x,n), q )  \in   \mathcal{F}_1, \mathcal{F}_2, \mathcal{F}_3  \}$
and $| E | = \ell$. For each $\mathcal{F}_i$, construct an $\ell$ dimensional vector 
by applying to each element of $E$ the function 
$$
f(p,x,q) = 
\left\{
\begin{array}{ll}
n & (p,(x,n),q) \in \mathcal{F}_i\\[2pt]
0 & \textit{otherwise.}
\end{array}
\right.\\
$$
$\square$

Such a comparison of dynamic behaviour results in  a spectrum of observations: 
\begin{itemize}
\item[($i.$)]  If the flux graphs are isomorphic,   
the differences in their edge weights will determine how similar they are,
whereby identical flux graphs will have a distance of 0.  
We refer to isomorphism here, rather than simple equivalence, 
because the compared graphs might have different  
sets of vertices, for example, due to different naming.  
\item[($ii.$)] if the two flux graphs have overlapping edges, but they 
are not isomorphic, then their uncommon edges 
will increase their distance. 
\item[($iii.$)] If the flux graphs are completely different, every edge of 
the two flux graphs will contribute to their distance. 
\end{itemize}

Below, we illustrate  our method on networks with varying sizes and connectivities from the literature, 
whereby we argue that the networks separated with a smaller distance produce more similar behaviours.

\section{Quantifying noise and comparing different time intervals of a CRN}
\begin{figure}[!b]
$$
\begin{array}{ll}
\begin{array}{l}
{\footnotesize\verb|r01: scupa + plg ->|}\\
{\footnotesize\verb|          pls + scupa  , 0.00006;|}\\
{\footnotesize\verb|r02: pls + scupa ->|}\\ 
{\footnotesize\verb|          tcupa + pls. , 0.0664;|}\\
{\footnotesize\verb|r03: tcupa + plg ->|}\\ 
{\footnotesize\verb|          pls + tcupa  , 0.0015;|}\\
{\footnotesize\verb|r04: scupa   ->         , 0.084;|}\\                           
{\footnotesize\verb|r05: plg     ->         , 0.032;|}\\                             
{\footnotesize\verb|r06: pls     ->         , 0.084;|}\\                             
{\footnotesize\verb|r07: tcupa   ->         , 0.084;|}\\                           
{\footnotesize\verb|r08: pls + X -> Xpls    , 50.0;|}\\                
{\footnotesize\verb|r09: Xpls    -> X + pls , 0.016;|}\\                 
{\footnotesize\verb|r10: Xpls    -> pls.    , 0.02;|}\\                          
{\footnotesize\verb|r11: X       ->         , 0.032;|}\\                               
{\footnotesize\verb|r12: Xpls    ->         , 0.032;|}\\                            
{\footnotesize\verb|r13:         -> scupa   , 1.9264;|}\\                            
{\footnotesize\verb|r14:         -> plg     , 6.02;|}\\                               
{\footnotesize\verb|r15:         -> X       , 6.02;|}                             
\end{array}
&
\begin{array}{cc}
\!\!\!\!
\!\!\!\!\includegraphics[width=45mm]{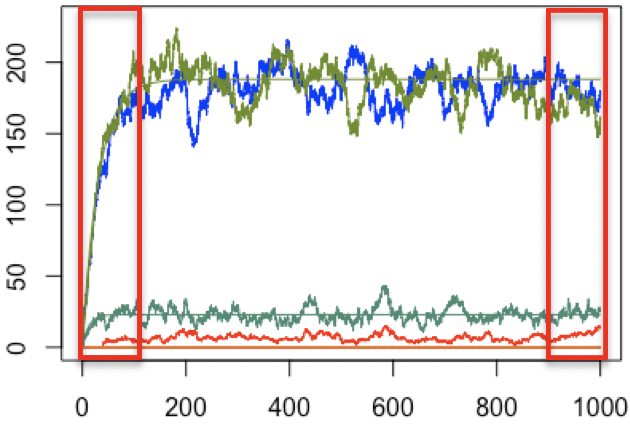}
\;\; & \;\;
\includegraphics[width=45mm]{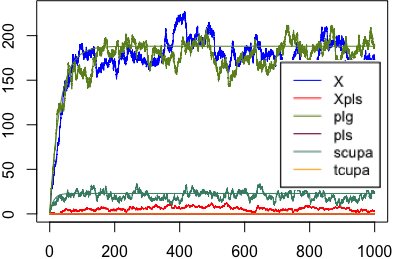}\\
\includegraphics[width=42mm]{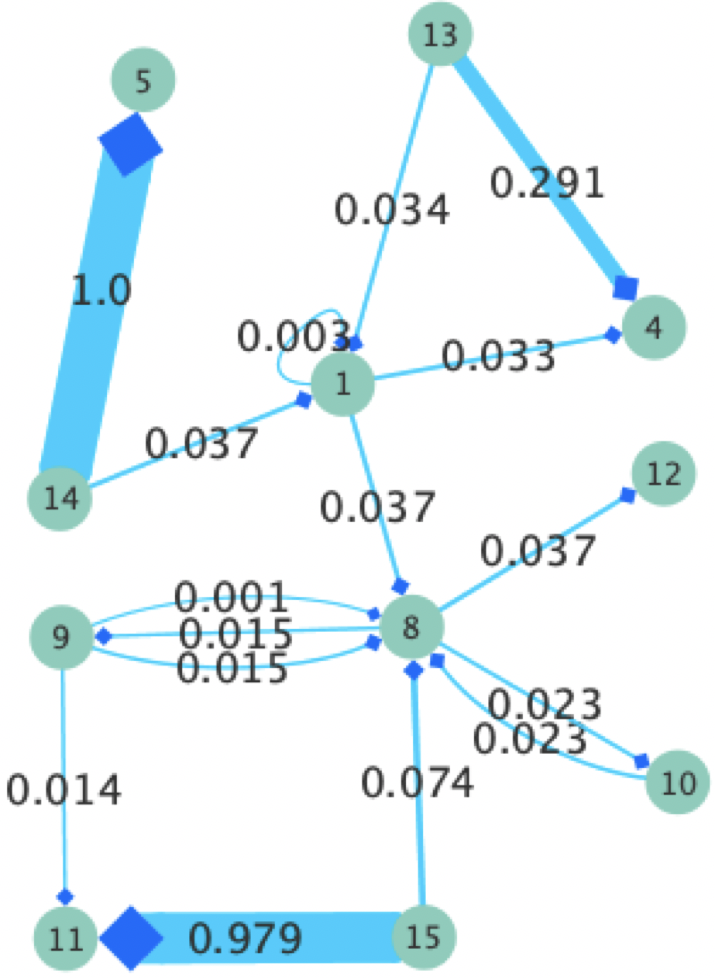}
\;\; & \;\; \;\;
\includegraphics[width=42mm]{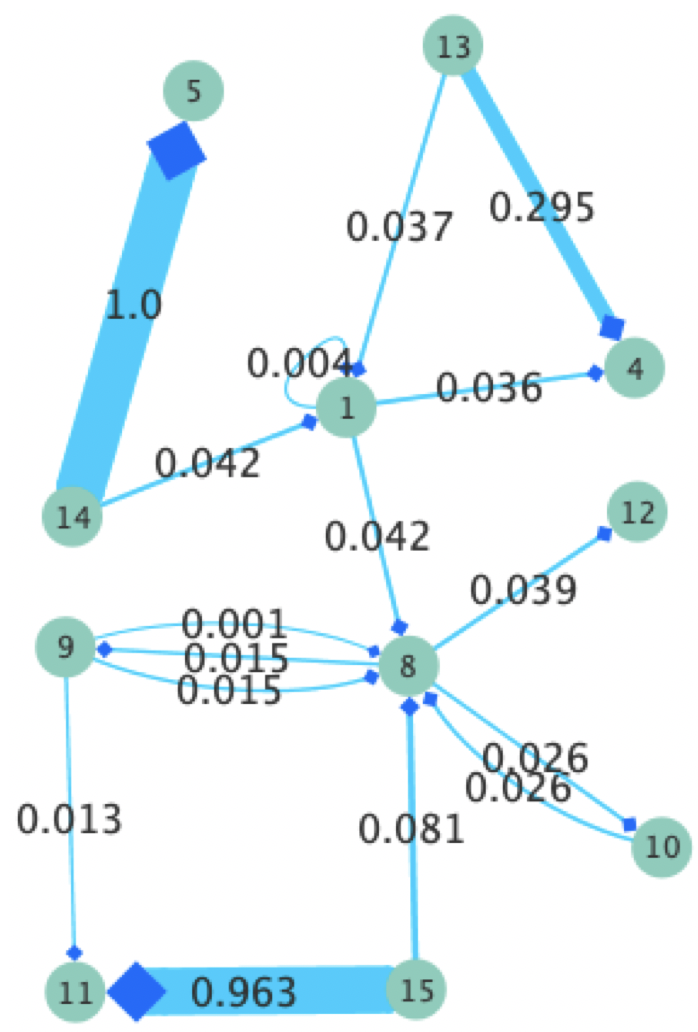}
\end{array}
\end{array}
$$
\caption{The CRN in  \cite{VLD11} that models  the  interactions of plasmin (\texttt{pls}) 
and urokinase-type plasminogen activator (\texttt{upa}). The two representative 
plots display the differences in stochastic time-series trajectories.
For comparison, the stochastic simulations are plotted together with the deterministic 
trajectory. 
The transient interval within the first 100 time units, 
whereby the system approaches the steady state, and the steady state interval 
between the time units 900 and 1000 are framed.   
The normalised flux graph of these two simulations are depicted below their 
time series.  The flux graphs are rendered with Cytoscape \cite{Sha03}.}
\label{figure:example:urokinase}
\end{figure}

The notion of distance between two flux graphs can be used to compare simulations of 
the same CRN to estimate the noise in the system.
As an example for this let us consider the CRN depicted in Figure \ref{figure:example:urokinase}, 
which models  the dynamic interactions of plasmin (\texttt{pls}) and urokinase-type plasminogen 
activator (\texttt{upa})  \cite{VLD11}. 
This network does not have any species instances at the initial state as it contains the  zero-order reactions
\texttt{r13}, \texttt{r14} and \texttt{r15} that introduce into the system the urokinase (\texttt{tcupa}), 
plasminogen (\texttt{plg}) and the generic substrate species \texttt{X}. The metabolites
\texttt{pls} and \texttt{upa} activate each other by interacting 
in their active and inactive forms. These interactions are modelled by the reactions  \texttt{r1}, \texttt{r2} and \texttt{r3}.
The reactions \texttt{r4}, \texttt{r5}, \texttt{r6}, \texttt{r7}, \texttt{r8} and \texttt{r11} model the 
degradation and dilution of the metabolites. 
The complexation and catalysis of various substrates by \texttt{pls} 
are modelled by the reactions    \texttt{r8}, \texttt{r9} and \texttt{r12}.

Figure \ref{figure:example:urokinase} displays the time series of two 
representative stochastic simulations with this network as well as their normalised flux graphs.
For comparison, the stochastic time series are plotted together with the deterministic trajectory of the system. 
 These two stochastic simulations demonstrate similar qualitative behaviours, 
 however they differ in their quantitative behaviour  due to the noise in the system. 
 To quantify the difference that these fluctuations create, we have computed the distance of 
 the normalised flux graphs in these two simulations. 
 The resulting distance value of 0.0334 provides an estimate 
 of the effect of noise that makes two such simulations different. 
 We have then performed this procedure on 100 pairs of simulations, 
and obtained a mean distance of $0.0315$ 
with a standard deviation of $0.01497$.

By taking this  estimated noise value as the base-line,
we have compared the dynamic behaviour of the system in the transient interval 
at the first 100 time units with the steady state interval in the last 100 time units, 
which can be observed in the deterministic trajectory.
These two intervals are depicted in frames in the time series of the first simulation in Figure 
\ref{figure:example:urokinase}.
Over a sample of 100 simulations the mean distance between these two intervals is  $0.1396$
with a standard deviation of $0.0398$. This distance value is greater than  the estimated 
noise by a factor of $4.43$ on a sample of same size, 
and thus indicates the extent to which the dynamic behaviour 
at the transient interval differs from the equilibrium behaviour.

\section{Quantifying information loss in model simplification}
 
\begin{figure}[!b]
$$
\begin{array}{l}
\!\!\!\!\!
\begin{array}{l}
\footnotesize{\verb|reactions|}\\
\footnotesize{\verb|r01: A + R  -> RA     , 1.0;|}\\
\footnotesize{\verb|r02: A + RD -> RDA    , 1.0;|}\\
\footnotesize{\verb|r03: A + RT -> RTA    , 1.0;|}\\
\footnotesize{\verb|r04: RA     -> A + R  , 500;|}\\  
\footnotesize{\verb|r05: RDA    -> A + RD , 500;|}\\ 
\footnotesize{\verb|r06: RTA    -> A + RT , 3.0;|}\\
\footnotesize{\verb|r07: E + R  -> RE     , 0.43;|}\\  
\footnotesize{\verb|r08: E + RD -> RDE    , 0.0054;|}\\ 
\footnotesize{\verb|r09: E + RT -> RTE    , 0.0075;|}\\ 
\footnotesize{\verb|r10: RE     -> E + R  , 1.074;|}\\ 
\end{array}
\qquad
\begin{array}{l}
\\ 
\footnotesize{\verb|r11: RDE -> E + RD , 0.136;|}\\  
\footnotesize{\verb|r12: RTE -> E + RT , 76.8;|}\\ 
\footnotesize{\verb|r13: RTA -> RDA    , 2104;|}\\ 
\footnotesize{\verb|r14: RDA -> RA     , 0.02;|}\\ 
\footnotesize{\verb|r15: RA  -> RDA    , 5.0;|}\\  
\footnotesize{\verb|r16: RA  -> RTA    , 4.25;|}\\  
\footnotesize{\verb|r17: RTA -> RA     , 0.0002;|}\\ 
\footnotesize{\verb|r18: RT  -> RD     , 0.02;|}\\ 
\footnotesize{\verb|r19: RD  -> R      , 0.02;|}\\  
\footnotesize{\verb|r20: R   -> RD     , 1.65;|}\\   
\end{array}
\qquad
\begin{array}{l}
\\
\footnotesize{\verb|r21: R   -> RT  , 50.0;|}\\ 
\footnotesize{\verb|r22: RT  -> R   , 0.02;|}\\ 
\footnotesize{\verb|r23: RTE -> RDE , 0.02;|}\\  
\footnotesize{\verb|r24: RDE -> RE  , 6.0;|}\\  
\footnotesize{\verb|r25: RE  -> RDE , 1.65;|}\\   
\footnotesize{\verb|r26: RE  -> RTE , 50;|}\\   
\footnotesize{\verb|r27: RTE -> RE  , 0.02;|}\\
\\
\footnotesize{\verb|initial state|}\\
\footnotesize{\verb|1000 R; 776 E; 10 A;|}\\
\end{array}
\\[80pt]
\begin{array}{cc}
\begin{array}{c}
\includegraphics[width=50mm]{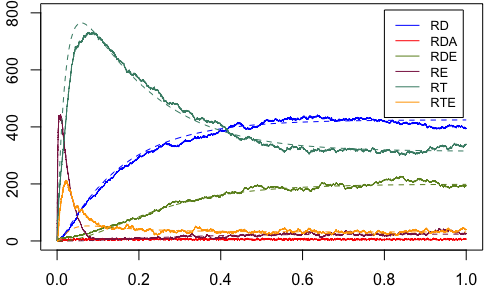}\\[2pt]
\includegraphics[width=50mm]{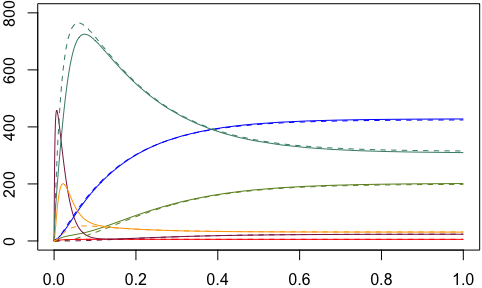}
\end{array}
&
\begin{array}{c}
\includegraphics[width=100mm]{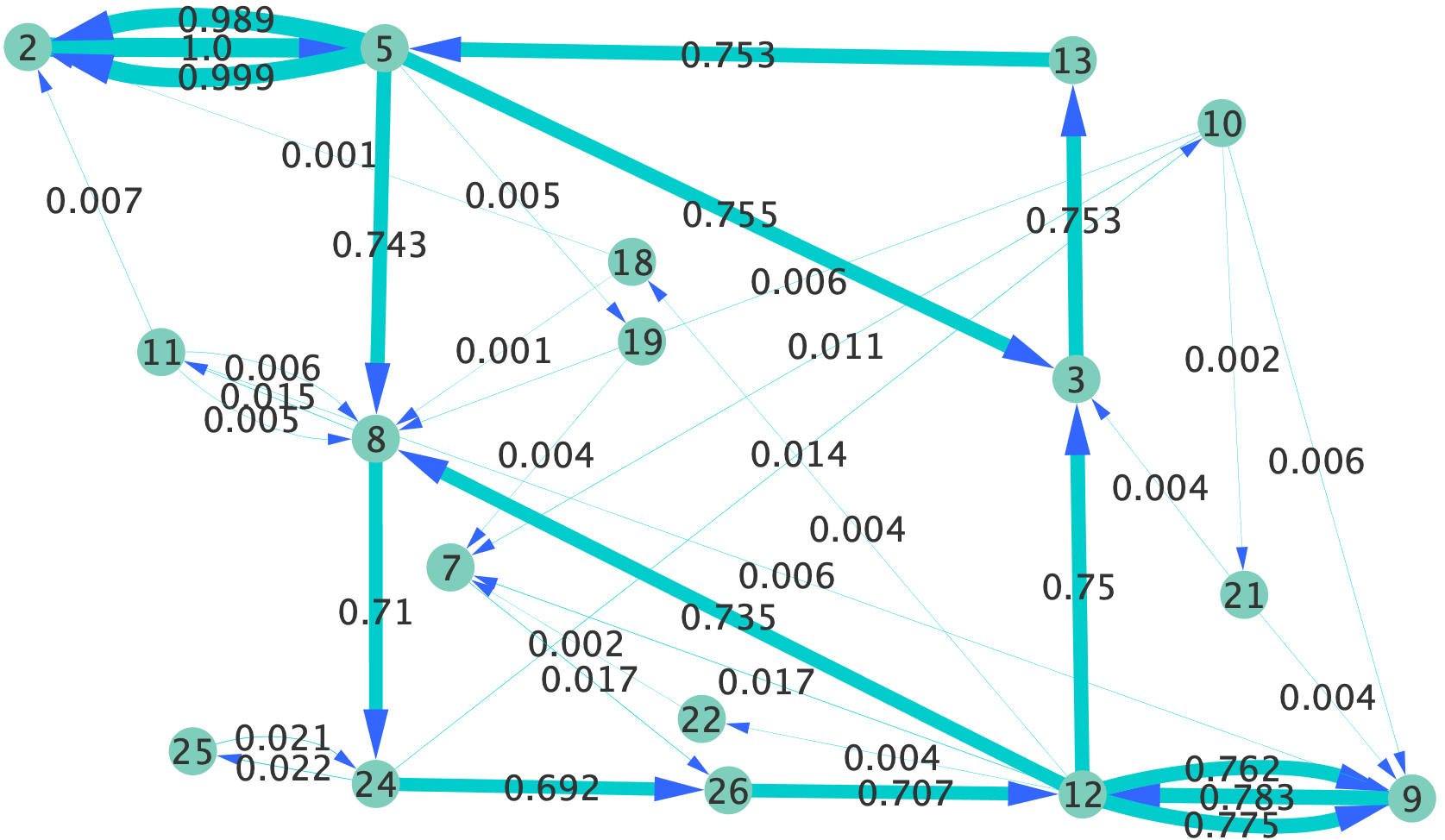}
\end{array}
\end{array}
\end{array}
$$
\caption{The CRN that models the Rho GTP-binding proteins and its flux graph \cite{CCGKP08,flux}.
In the time series on the top, a stochastic simulation trajectory with this network 
is plotted together with the deterministic ordinary differential equations trajectory, which is 
indicated with the dashed lines. The reactions 2, 3, 5, 8, 9, 12, 13, 24, 26
 that participate in the strong fluxes in the flux graph and the reaction 21 that introduces 
 the active metabolite \texttt{RT}  are used to construct a simplified network.
 The time series below displays the  deterministic time series with the full network
 together with the deterministic time series of the reduced network, plotted with the dashed lines.
The flux graph is rendered with Cytoscape \cite{Sha03}.}
\label{figure:example:rho}
\end{figure}

In systems biology, it is often desirable to replace a model with a simpler 
one that produces a similar dynamics with the minimum loss of information.
As an example for this, let us consider the CRN in Figure  \ref{figure:example:rho}
that models the Rho GRP binding proteins that cycle between inactive (\texttt{RD})
and active (\texttt{RT}) forms by being catalysed by the enzymes \texttt{A} and \texttt{E}.
In previous work \cite{flux}, we have shown that flux analysis can be used to simplify 
this network to its subset that consists of the reactions participating in strong flux edges.     
Such a simplified network, consisting of the reactions 2, 3, 5, 8, 9, 12, 13, 21, 24 and 26,
produces a similar time series as shown in Figure \ref{figure:example:rho}. 
Here, although reaction 21 does not participate in a strong flux, 
it is included in the simplified network 
as it introduces the active metabolite \texttt{RT}. 

To quantify the loss of information in the simplification, we have first computed the 
distance between the simulations of the full network to obtain an estimate of noise. 
On a sample of 100 pairs of simulations, the mean distance between two 
simulations of the full network is $0.1135$
with a standard deviation of $0.05723$. The distance between the complete network and the simplified 
network for the full interval, on the other hand, accounts for a distance of $1.13$ on a sample of 100 simulations.  
When we compare the transient interval up to the time point 0.5, this distance becomes $1.74$. 
However, the distance in the steady state interval between the time points 0.5 and 1.0, 
again on a sample of 100 simulations, is $0.0987$, which is in the order of noise base-line.

These results demonstrate that for the steady state interval 
the simplified network's information loss is in the order of noise base-line. 
Moreover, the complete loss of information in the simplified network is due to the transient 
interval, whereby the system species stabilise while approaching equilibrium. 

\section{Quantifying variation in behaviour in different regimes of a network}

 
\begin{figure}[!b]
$$
\begin{array}{l}
\includegraphics[width=160mm]{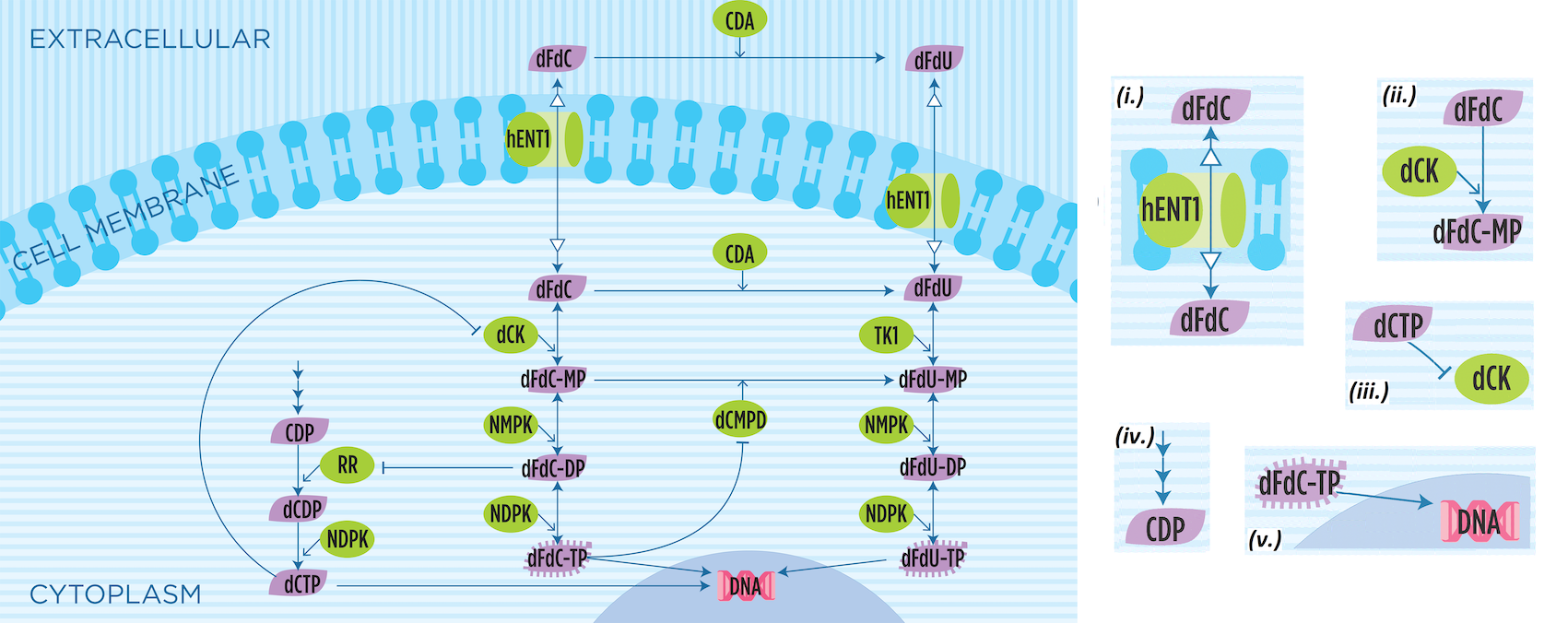}\\[10pt]
\begin{array}{c}
\!\!\!\!\!\!\!\!\!\!\!\!
\includegraphics[width=68mm]{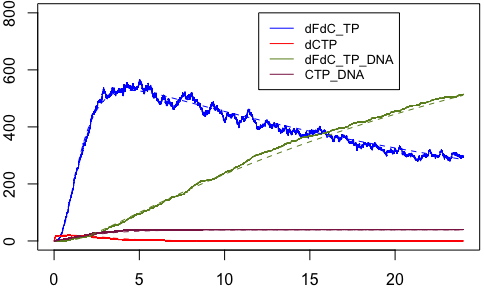}
\end{array}
\;
\begin{array}{c}
\includegraphics[width=1.5mm]{Images/separator.png}
\end{array}
\;
\begin{array}{c}
\includegraphics[width=68mm]{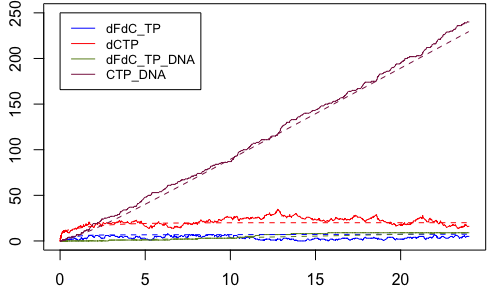}
\end{array}
\\
\!\!\!\!\!\!\!\!\!\!\!\!
\begin{array}{c}
\includegraphics[width=68mm]{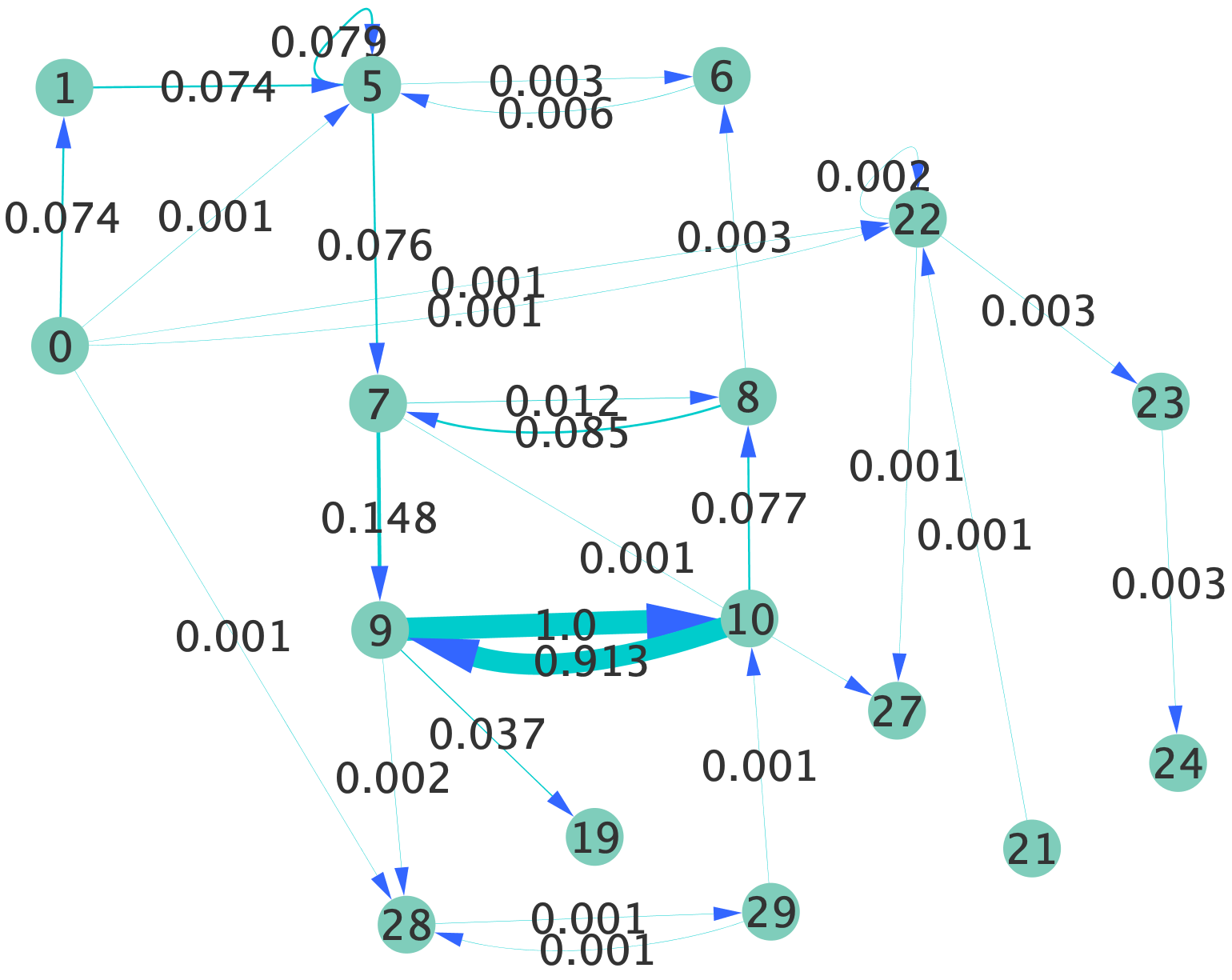}
\end{array}
\;
\begin{array}{c}
\includegraphics[width=1.5mm]{Images/separator.png}
\end{array}
\;
\begin{array}{c}
\includegraphics[width=82mm]{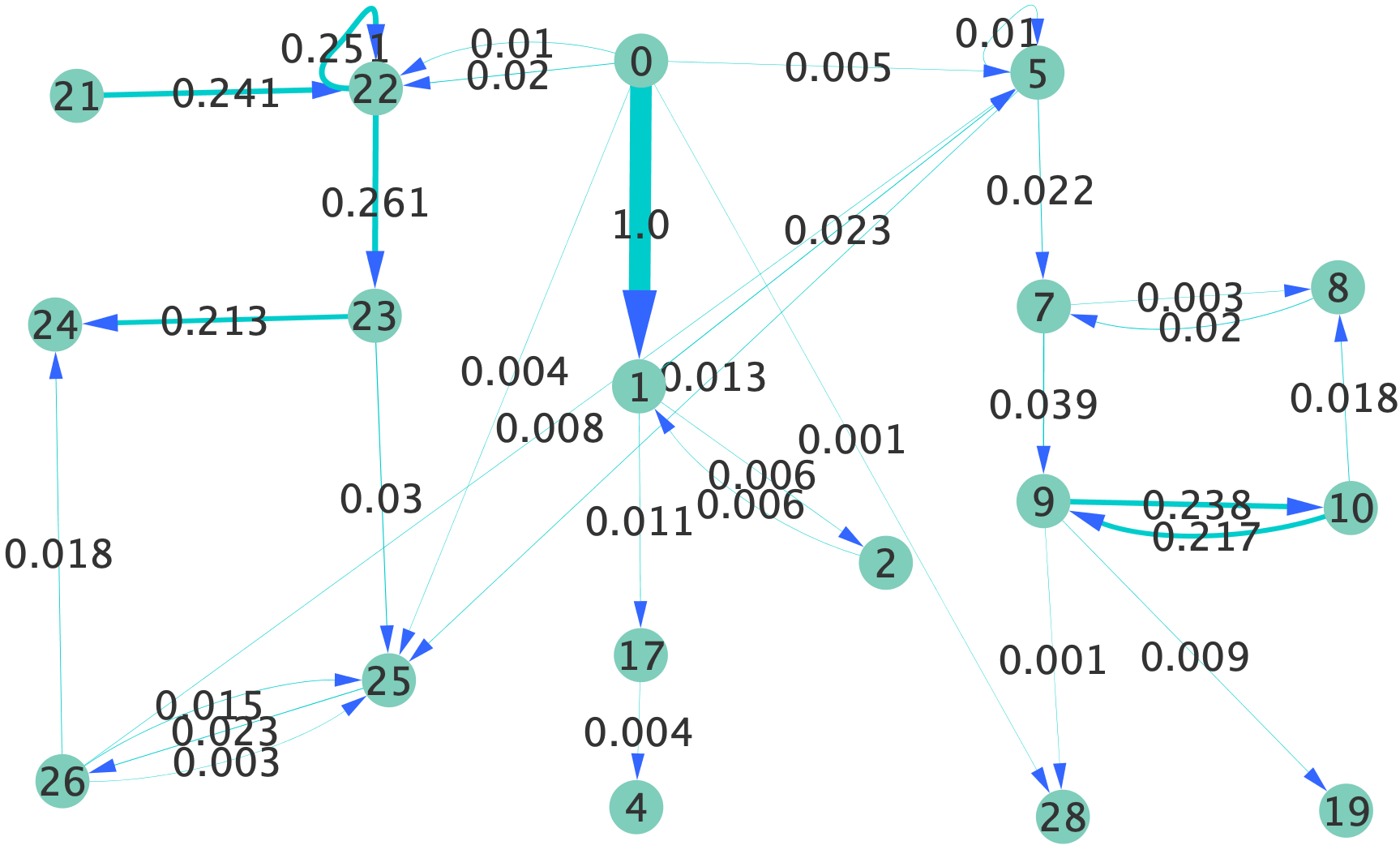}
\end{array}
\end{array}
$$
\caption{The molecular machinery of the Gemcitabine metabolism, 
representative time series plots in the extremities of the efficacy/non-efficacy continuum, 
and the mean normalised flux graphs on a sample of 100 simulations in these regimes \cite{Kah12}. 
The arrows on the right describe ($i.$) the transport into the cell, ($ii.$) enzymatic reactions, 
($iii.$) inhibition, ($iv.$) synthesis,  and ($v.$) DNA incorporation.
The corresponding CRN is shown in Figure \ref{figure:example:gemcitabine:CRN}.
The time series plot on the left for the efficacy regime is obtained from a single simulation
with $\alpha = 0.000001$ and $\beta = 0.01$. The one on the right for the non-efficacy 
regime is obtained from a single simulation  with $\alpha = 10$ and $\beta = 0.000001$.
For comparison, both plots include the deterministic ordinary equation 
trajectories indicated with dashed lines.
The flux graphs of the efficacy regime on the left and 
non-efficacy regimes on the right are rendered with Cytoscape \cite{Sha03}.}
\label{figure:gemcitabine:cartoon}
\end{figure}



\begin{figure}[!b]
{\verb|reactions|}\\ 
{\verb|r01: dFdC_out           -> dFdC               , 9.97234;|}  \\
{\verb|r02: dFdC               -> dFdC_out           , 0.000261675;|}\\
{\verb|r03: dFdC_out           -> dFdU               , 0.00000472336;|}\\
{\verb|r04: dFdU               -> dFdU_out           , 0.0508194;|}\\
{\verb|r05: dFdC + dCK         -> dFdC_MP + dCK      , 0.104994;|}\\
{\verb|r06: dFdC_MP            -> dFdC               , 0.0875208;  |}\\
{\verb|r07: dFdC_MP            -> dFdC_DP            , 2.37162;  |}\\
{\verb|r08: dFdC_DP            -> dFdC_MP            , 0.212216;  |}\\
{\verb|r09: dFdC_DP            -> dFdC_TP            , 2.52037;  |}\\
{\verb|r10: dFdC_TP            -> dFdC_DP            , 1.44908;|}\\
{\verb|r11: dFdU + dCK         -> dFdU_MP + dCK      , 0.0968;  |}\\
{\verb|r12: dFdU_MP            -> dFdU               , 0.0000560415; |}\\    
{\verb|r13: dFdU_MP            -> dFdU_DP            , 0.07844;    |}\\
{\verb|r14: dFdU_DP            -> dFdU_MP            , 0.00420541;   |}\\ 
{\verb|r15: dFdU_DP            -> dFdU_TP            , 0.164322;     |}\\ 
{\verb|r16: dFdU_TP            -> dFdU_DP            , 0.0000905139; |}\\   
{\verb|r17: dFdC               -> dFdU               , 0.000476746;  |}\\  
{\verb|r18: dFdC_MP + dCMPD    -> dFdU_MP + dCMPD    , 0.0004559;    |}\\
{\verb|r19: dFdC_TP            -> dFdC_TP_DNA        , 0.0544456;    |}\\
{\verb|r20: dFdU_TP            -> dFdU_TP_DNA        , 0.000737496;  |}\\
{\verb|r21:                    -> CDP                , 10;    |}\\
{\verb|r22: CDP + RR           -> dCDP + RR          , 5;    |}\\
{\verb|r23: dCDP               -> dCTP               , 25.2037;    |}\\
{\verb|r24: dCTP               -> CTP_DNA            , 0.5;|}\\
{\verb|r25: dCTP + dCK         -> dCTP_dCK           , alpha;|}\\ 
{\verb|// efficacy: alpha = 0.000001 ~ non-efficacy: alpha = 10|}\\  
{\verb||}\\    
{\verb|r26: dCTP_dCK           -> dCTP + dCK         , 0.1;  |}\\
{\verb|r27: dFdC_DP + RR       -> dFdC_DP_RR         , beta;|}\\
{\verb|// efficacy: beta = 0.01 ~ non-efficacy: beta = 0.000001  |}\\
{\verb||}\\
{\verb|r28: dFdC_TP + dCMPD    -> dFdC_TP_dCMPD      , 0.001;|}\\
{\verb|r29: dFdC_TP_dCMPD      -> dFdC_TP + dCMPD    , 0.1; |}\\
\\
{\verb|initial state|}\\
{\verb|1000 dFdC_out; 10 dCK; |} 
{\verb|10 dCMPD;  20 CDP;  10 RR;|}\\
\caption{The CRN that implements the model depicted in Figure \ref{figure:gemcitabine:cartoon}.}
\label{figure:example:gemcitabine:CRN}
\end{figure}

A CRN with a certain set of reactions can have drastically different time series 
behaviours with variations in initial conditions or rate constants.  
Because such parameters are used to represent 
metabolic conditions in different regimes,
quantifying the deviation in behaviour that they cause can provide insights 
 for the modelled biological systems. An application of this idea is illustrated in 
 Example \ref{example:sensitivity:anlaysis} and  
 Figure \ref{figure:sir:scatter:plot}.

 Let us now consider a larger CRN, depicted in Figures \ref{figure:gemcitabine:cartoon} 
 and \ref{figure:example:gemcitabine:CRN}, that models the 
 Gemcitabine biochemical machinery  \cite{Kah12}. 
 Gemcitabine is a prodrug, 
 which is widely used for treating various carcinomas. 
Gemcitabine 
(\texttt{dFdC}) exerts its clinical effect by incorporating its 
triphosphate metabolite (\texttt{dFdC-TP}) 
into DNA, thereby inhibiting DNA synthesis and replication, 
eventually resulting in programmed cell death.   
This process takes place in competition with the natural nucleoside \texttt{dCTP}. 
Besides the indirect competition for incorporation into DNA, 
\texttt{dFdC-TP} and \texttt{dCTP}
compete by direct competition, whereby they 
inhibit the enzymes in each other's activation cascades.  
The relative abundance of these enzymes, \texttt{RR}  and \texttt{dCK}, 
gives rise to an efficacy/non-efficacy spectrum of the drug on a continuum. 
In the CRN in Figure \ref{figure:example:gemcitabine:CRN},
this continuum is modelled by changes in the rate constants $\alpha$ and $\beta$
of the reactions 25 and 27. On one extreme,  the efficacy regime is 
given by $\alpha = 0.000001$ and $\beta = 0.01$. On the other extreme, 
the non-efficacy regime is given by $\alpha = 10$ and $\beta = 0.000001$.
These two regimes result in drastically different time series for 
the metabolites \texttt{dFdC\_TP\_DNA} and \texttt{dCTP\_DNA} 
that determine the efficacy of the drug.

To estimate the behavioural difference between the efficacy 
and non-efficacy regimes, 
we have first computed the mean distance of the normalised 
flux graphs 
within these regimes as the base-line. 
On a sample of 100 pairs of simulations, the mean distance 
between two simulations in the efficacy regime 
is $0.0082$ with a standard deviation of $0.0035$. 
In the non-efficacy regime, 
the distance is $0.1364$ with a standard deviation of $0.081$. 
On this sample of 100 simulations, the mean distance 
between the efficacy and non-efficacy regimes is 1.48. 
When we take the efficacy regime as the base-line for comparing 
the efficacy and non-efficacy regimes, we obtain 
the ratio $1.48/0.1364$ that delivers a factor of $10.85$. 
Taking the non-efficacy regime as the base-line 
results in the ratio of $1.48/0.0082$, delivering a factor of $180.48$.  
This allows us to use this later factor as a quantification of 
the upper estimate for the variation between the two regimes 
at the extremities of the efficacy/non-efficacy spectrum.

\section{Discussion}

CRNs provide a convenient and expressive 
programming language for modelling biological systems as well as other 
complex systems across many branches of science. Besides their role as back-end
engines in rule-based modelling languages \cite{BFKF18,SF12}, 
their use in synthetic biology in implementing molecular programs 
resembles common programming practises  \cite{LYPEP11,Kah15}.  
However, the quantitative nature of CRNs, which heavily relies on parameterisation, 
makes them difficult to analyse and compare in a qualitative manner. This is because,
unlike programs in common programming languages,
the complex behaviours that emerge from CRN runs, i.e., simulations,   
often remain undetectable from their structure. 

Despite the inherent challenges, the methods that originate from computer science, 
in particular those in concurrency theory, have been studied by various authors with the aim of  
connecting network structure with quantitative behaviour. 
These efforts mainly rely on the consideration that the analogy between 
complex systems studied in computer science 
and those in other disciplines  is prone to providing new insights 
that can benefit both sides.  
In this respect,  the flux graphs of our method has its origins in concurrency theory 
due to the resemblance between event structures \cite{NPW81,Kah09} 
and Markov chain semantics of CRNs: the dependency graphs (Figure \ref{figure:trajectory:to:flux}), 
which are precursors 
of the flux graphs, can be treated as configurations in event structures.
Such a view of these structures as well as those from other branches of computer science 
can provide further perspectives.  In this regard, 
an approach that also uses partial order representations of traces for 
comparing concurrent programs can be found in \cite{BEL17}. 
Similar considerations that use distance metrics on various aspects of
programs  can  provide insights  in comparing programs in a quantitative 
manner for the cases, where a qualitative comparison is not straight-forward or meaningful.

In quantitative disciplines, being able compare the primary structures 
of function, possibly in a way that permits for a notion of equivalence, is a 
key to evaluate and gain insight on such structures.  
Our method  for comparing CRNs with respect to their dynamic 
behaviours in simulations is set out with such an ambition. 
In the context of comparing CRNs, a promising method that is orthogonal to the one 
proposed here and combines quantitative 
and structural points of views is given in \cite{CTTV17}, 
where the authors describe a method, with certain constraints, 
for minimising a CRN by aggregating its species.
The notion of species aggregation as well  as reaction aggregation 
is a topic of ongoing investigation also for the present work. 
In particular, such aggregations can provide further aid  when comparing very large networks 
or comparing the behaviour of a large network with a smaller one.
 Another related topic, which is left for future work, 
concerns algorithms for re-mapping reaction nodes of compared flux graphs. 
This is because two networks that appear different at a first observation 
can be separated by a small distance after renaming their reaction edges, 
and this can explain the similarity in their dynamic behaviours.  

Our method uses flux graphs resulting from  
stochastic simulations as discrete abstract 
representations of dynamic behaviour. 
This makes it possible to quantify  
the simulations in terms of their distance at steady states as well as 
at transient time intervals at which the system species have 
not yet reached equilibrium. The use of a metric in our method 
makes it possible to assign a distance measure to the comparisons  
that overlaps with graph isomorphism at the closest distance. 
Because flux graphs have intuitive interpretations as they 
display which portion of simulation species 
flows between which reactions, 
they constitute favourable abstractions of behaviour that
preserve the inherent stochasticity of the CRNs. 
Our comparison of flux graphs uses the notion that graphs that 
are close according to our metric have similar intensities of fluxes.
A deeper analysis on when and why two graphs are close 
according to the metric is left for future work.

Due to the underlying mass action semantics, the observations obtained 
with our method complement those that are obtained with
deterministic simulations with ordinary differential equations.
Although differential equations do not directly capture 
noise and stochasticity, in some cases 
methods based on linear noise approximation \cite{CKL16} 
can be used for recovering the information on stochasticity.    
Still, the discrete structure of flux graphs makes them favourable 
structures to work with algorithmically~\footnote{The python
scripts and tools of our method are available for 
download at \texttt{ozan-k.com/distance.zip}.}, 
and their counter-part in the 
continuous setting of ordinary differential equations is yet to be identified.

\bibliographystyle{eptcs}
\bibliography{ozan}
\end{document}